\documentclass[aps,prb,amsfonts,amssymb,notitlepage,tightenlines,twocolumn,showpacs]{revtex4-1}
\usepackage[intlimits]{amsmath}
\usepackage{graphicx}

\usepackage{physics}

\usepackage{color}

\newcommand{\rr}{\mathbf{r}}

\newcommand{\HH}{\mathcal{H}}

\newcommand{\bosi}{\boldsymbol{\sigma}}

\begin{document}

\title{Electronic and Optical Properties of Vacancy Defects in Transition Metal Dichalcogenides}

\author{M. A. Khan$^{1,2}$}
\author{Mikhail Erementchouk$^{1,3}$}
\author{Joshua Hendrickson$^4$}
\author{Michael N. Leuenberger$^1$}
\affiliation{$^{1}$NanoScience Technology Center, Department of Physics, and College of Optics and Photonics, University
of Central Florida, Orlando, FL 32826, USA}
\affiliation{$^{2}$Department of Aplied Physics, Federal Urdu University of Arts, Science and Technology, Islamabad, Pakistan}
\affiliation{$^{3}$Department of Electrical Engineering and Computer Science, University of Michigan, Ann Arbor,
MI 48109, USA}
\affiliation{$^{4}$Air Force Research Laboratory, Sensors Directorate, Wright-Patterson Air Force Base, Ohio, 45433, USA}

\begin{abstract}
A detailed first-principle study has been performed to evaluate the electronic and optical properties of single-layer (SL) transition metal dichalcogenides (TMDCs) (MX${}_2$; M= transition metal such as Mo, W and X= S, Se, Te), in the presence of vacancy defects (VDs). Defects usually play an important role in tailoring electronic, optical, and magnetic properties of semiconductors. We consider three types of VDs in SL TMDCs i) $X$-vacancy, $X_{2}$-vacancy, and iii) $M$-vacancy. We show that VDs lead to localized defect states (LDS) in the band structure, which in turn give rise to sharp transitions in in-plane and out-of-plane optical susceptibilities, $\chi_{\parallel}$ and $\chi_{\perp}$. The effects of spin orbit coupling (SOC) are also considered. We find that SOC splitting in LDS is directly related to the atomic number of the transition metal atoms. Apart from electronic and optical properties we also find magnetic signatures (local magnetic moment of $\sim\mu_{B}$) in MoSe$_{2}$ in the presence of Mo vacancy, which breaks the time reversal symmetry and therefore lifts the Kramers degeneracy. We show that a simple qualitative tight binding model (TBM), involving only the hopping between atoms surrounding the vacancy with an on-site SOC term, is sufficient to capture the essential features of LDS. In addition, the existence of the LDS can be understood from the solution of the 2D Dirac Hamiltonian by employing infinite mass boundary conditions. In order to provide a clear description of the optical absorption spectra, we use group theory to derive the optical selection rules between LDS for both $\chi_{\parallel}$ and $\chi_{\perp}$. 
\end{abstract}

\maketitle

\section{Introduction}

Single-layer (SL) transition metal dichalcogenides (TMDCs) have attracted a lot of attention due to their intriguing electronic and optical properties, with a wide range of promising applications.\cite{review_TMDCS, review_TMDCS_2} SL TMDCs are direct band gap semiconductors\cite{Direct_Band_Gap_1, Direct_Band_gap_2}, which can be used to produce smaller and more energy efficient devices, such as transistors and integrated circuits. Moreover, the band gap lies in the visible region which makes them highly responsive when exposed to visible light, a property with potential applications in optical detection. In contrast to graphene, SL TMDCs exhibit large intrinsic spin-orbit coupling (SOC), originating from the d orbitals of transition metal atoms. The presence of \textcolor{black}{considerably high} SOC (up to few hundred meV)\textcolor{black}{\cite{SOC_1,SOC_2,Song_SOI}} makes them a candidate material for exploring spin physics and spintronics applications.

Wafer-scale production of SL TMDCs is required to fully appreciate their technological potential.  The most common experimental techniques used to produce large chunks of SL MoS$_{2}$ are i) mechanical exfoliation, ii) chemical vapor deposition, and iii) physical vapor deposition. It has been observed that samples produced by all of these techniques have considerably lower carrier mobility than the theoretically predicted values.\cite{reduced_mobility_1, reduced_mobility_2} It has recently been suggested that this discrepancy between the predicted and observed values of carrier mobility is due to the presence of impurities created during the growth process.\cite{reduced_carrier_mobility_1, reduced_carrier_mobility_2} The most common and energetically favorable types of impurities are vacancy defects (VDs).\cite{Hong_2015}
\begin{figure*}[bt]
	\begin{center}
		\includegraphics[width=6.5in]{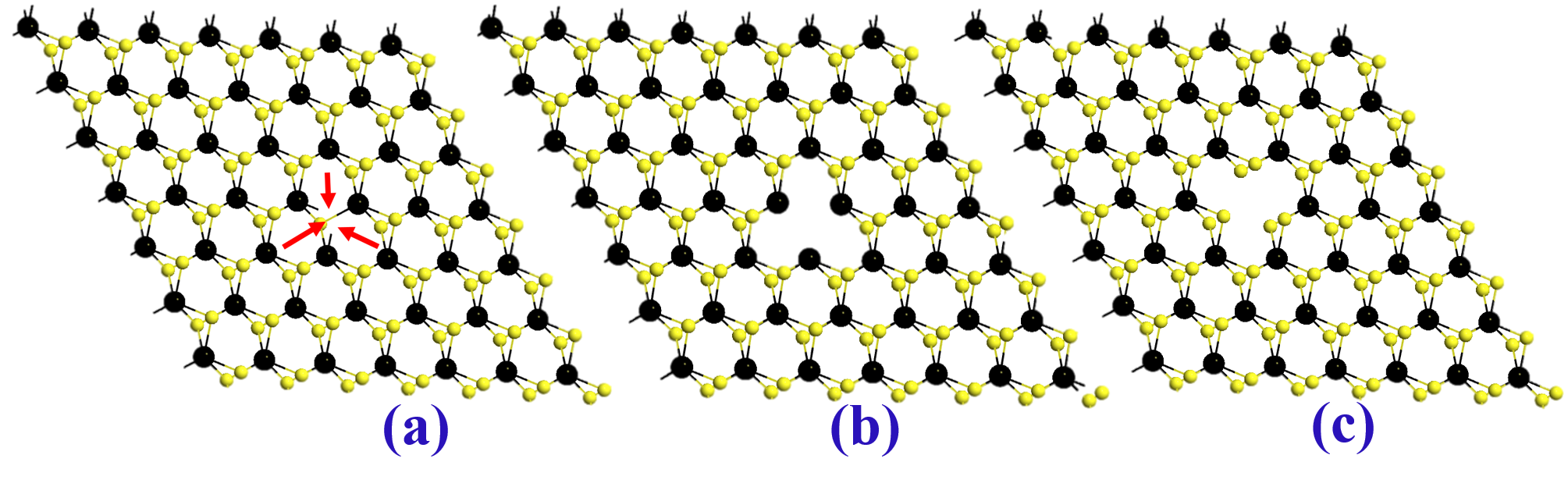}
	\end{center}
	\caption{Different types of vacancy defects. Yellow sphere is X while blue sphere is M (a) X-vacancy in $7 \times 7 \times 1$ supercell, \textcolor{black}{(b)} X$_2$-vacancy consisting of a pair of X atoms (surrounded by M atoms) removed in $7 \times 7 \times 1$ supercell, \textcolor{black}{(c)} M vacancy (surrounded by X atoms) in $7 \times 7 \times 1$ supercell.}
	\label{fig:structure_1}
\end{figure*}
Defects usually play an important role in tailoring various electronic and optical properties of two dimensional materials and have been the subject of intense research over the last few decades. VDs in semiconductors act as trapping centers for charge carriers and their interaction with charge carriers becomes stronger at reduced dimensionalities. \textcolor{black}{Point defects in SL TMDCs have been explored both theoretically and experimentally.\cite{Defect_Kunstmann,Defect_Pulkin,Defect_Li,sirivastav_SPS, Yu_Ming_He_SPS, KoperskiM_SPS}} Recent photoluminescence (PL) experiments \cite{sirivastav_SPS, Yu_Ming_He_SPS, KoperskiM_SPS} reveal that localized excitonic states related to VDs can serve as single-photon emitters in WSe$_{2}$. Magnetism in low dimensional systems is another area of interest.\cite{Grahene_Mag, Graphene_mag_yazyev} It has been shown that\cite{magnetic_Mo_MoSe2} certain LDS related to VDs can induce ferromagnetism in SL TMDCs, suggesting that they could be good candidates for spin channels in spintronic devices. In addition, LDS can be used to open and tune a band gap in graphene\cite{Graphene_Antidot} and SL MoS$_{2}$.\cite{Huang_2013, BG_Tune_MoS_2} Various atomic defects can be realized artificially by using different experimental techniques. It has been shown that hexagonal pits (3$\times$MoS$_{2}$) can be removed through etching of MoS$_{2}$ crystals by using XeF$_{2}$ as a gaseous reactant.\cite{Huang_2013} Point defects can be induced by irradiating the SL TMDCs with $\alpha$-particles or by thermal annealing.\cite{alphar-particle} \textcolor{black}{Several experimental studies have been reported regarding the effects of point defects or of grain boundaries on SL TMDCs.\cite{PL_MoS2, ARI_MoS2, EI_MoS2} Strong PL enhancement has been observed as a result of oxygen adsorption at sulfur vacancy sites.\cite{ARI_MoS2} Also, sulfur vacancies are observed in MoS$_{2}$ through transmission electron microscopy experiments.\cite{EI_MoS2}}
 
Pristine TMDCs are invariant with respect to the reflection $\sigma_{h}$ about the Mo or W plane of atoms ($z=0$ plane). Therefore, electron states can be classified into two catagories: even and odd or symmetric and antisymmetric with respect to the $z=0$ plane. We found that the even and odd bands in TMDCs have two different band gaps $E_{g\parallel}$ and $E_{g\perp}$, respectively.\cite{First_paper} $E_{g\perp}$ has been usually neglected for pristine TMDCs because of its substantially larger value (Table~\ref{Pristine_table}) and weak optical response (Fig:~\ref{fig:BS_pristine} b) as compared with $E_{g\parallel}$. Earlier studies\cite{First_paper,BG_Tune_MoS_2} show that the presence of VDs gives rise to  LDS  in addition to the normal extended states present in conduction or valence bands in SL MoS$_{2}$. These LDS appear within the band gap region and they can also be present deep inside the valence band depending on the type of VD. Optical transitions between LDS across Fermi level appear as resonance peaks, both in $\chi_{\parallel}$  and $\chi_{\perp}$, which shows that odd states are necessary for understanding the properties of VDs in SL MoS$_{2}$.\cite{First_paper} 

\textcolor{black}{In this paper, our aim is fourfold. First, we provide a comprehensive study of VDs in 4 types of SL TMDC materials, MoS$_2$, MoSe$_2$, WS$_2$, and WSe$_2$. Second, we provide detailed analytical models about the description of LDS within the Dirac equation formulation and by using the tight binding model.  Third, we include the effects of SOC on VDs, which has not been considered so far.  As mentioned earlier, SOC in these materials is large and therefore needs to be taken into account in order to obtain a better understanding of the electronic and optical properties of TMDCs. Fourth, we briefly discuss defect induced magnetism in some cases. Throughout this work, we consider 3 types of VDs:  i) Single $X$-vacancy ii) $X_{2}$-vacancy, and iii) $M$-vacancy.}

This paper is organized as follows. Section~\ref{sec:bandstructure} describes the numerical results obtained for band structures. Sections~\ref{sec:TBM} and \ref{sec:Dirac} describes qualitative models for the existence of defected states. Section~\ref{sec:optical} deals with the optical response of defected SL TMDCs. 

\section{Band Structure}
\label{sec:bandstructure}

\begin{figure*}[bt]
	\begin{center}
		\includegraphics[width=7in]{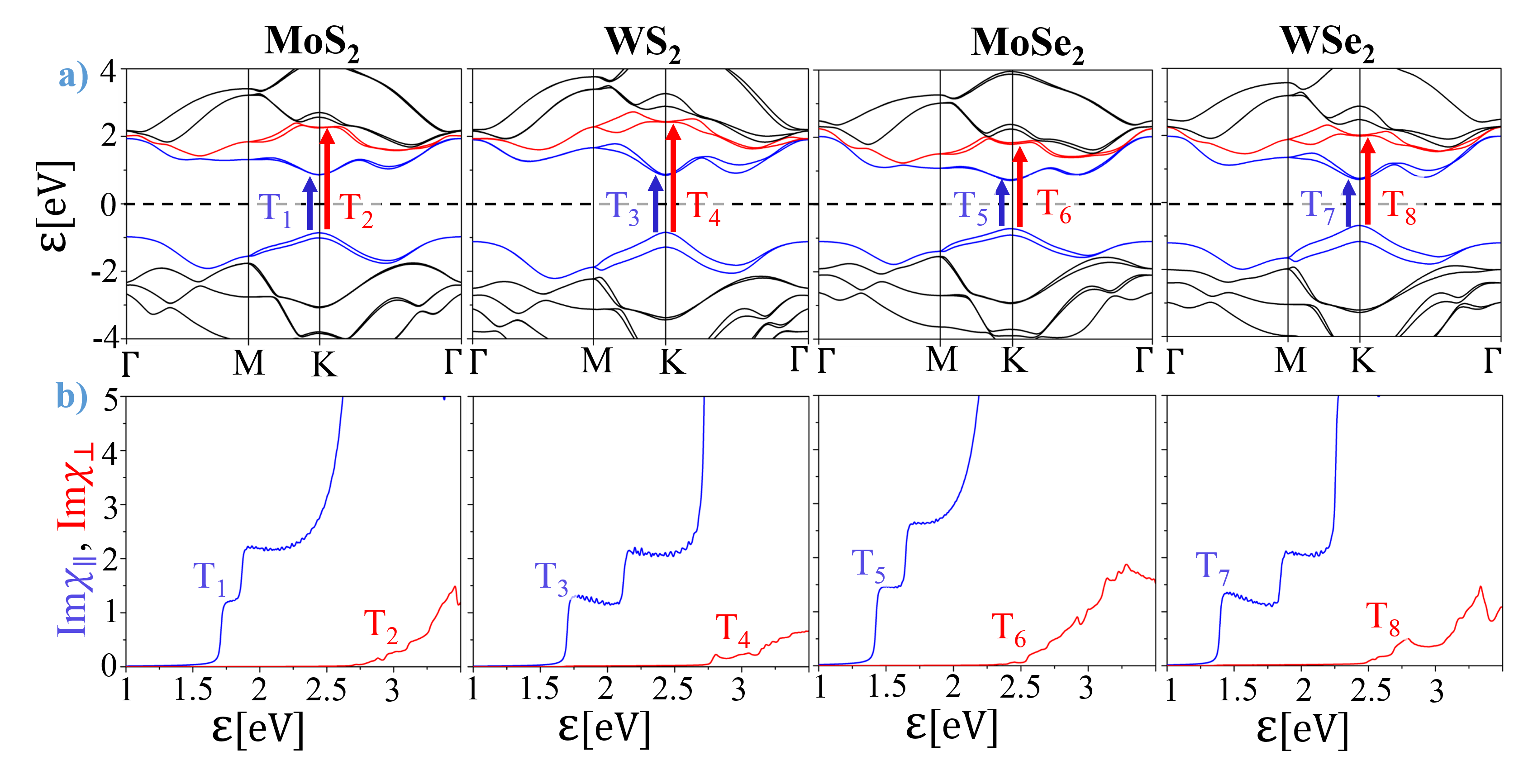}
	\end{center}
	\caption{(a) Band structures and (b) electrical susceptibility of pristine $MX_{2}$ monolayers. Band gaps $E_{g}^{\parallel}$ (blue), $E_{g}^{\perp}$ (red) and spin-orbit splitting $\Delta_{SO}$ in the valance band are given in Table~\ref{Pristine_table}. Spin splitting can also be seen in the diagonal electric susceptibility in the in-plane component $\mathrm{Im}(\chi_{\parallel})$.}
	\label{fig:BS_pristine}
\end{figure*}

The model system consists of a periodic 2D superlattice of TMDCs (Fig:~\ref{fig:structure_1} a, b, c). All numerical calculations are carried out using density functional theory (DFT). The local density approximation (LDA) is used  with the Perdew-Zunger (PZ) parametrization\cite{PZ_functionals} of the correlation energy of a homogeneous electron gas calculated by Ceperley-Alder \cite{Ceperley_Alder}. The calculations are implemented within Atomistix Toolkit 2015.1\cite{QW_1} \textcolor{black}{in order to be able to perform DFT calculations on large supercells in a reasonable amount of time}. The periodic structure of the superlattice allows one to characterize the electron states by the bandstructure $\epsilon_n(\mathbf{k})$, where $\mathbf{k}$ is the vector in the first Brillouin zone of the superlattice and $n$ enumerates different bands. We consider a $7 \times 7 \times 1$ (Fig:~\ref{fig:structure_1}) supercell having $147$ number of atoms with an edge length of $21.354$ \AA. The Brillouin zone of the supercell is sampled by a $3 \times 3 \times 1$ $k$-mesh. All the structures are geometrically optimized with a force tolerance of $0.05$ eV/\AA. SOC is taken into account via the norm conserving pseudo potentials. \cite{Norm_conserving_pseudopotentials_1,Norm_conserving_pseudopotentials_2} Band structures are calculated along the $\Gamma$ $-$ $M$ $-$ $K$ $-$ $\Gamma$ path. Band structures of SL TMDCs for the pristine cases are plotted in Fig.~\ref{fig:BS_pristine} and calculated values are given in Table~\ref{Pristine_table}. The results are in good agreement with previously reported values both for band gap and SOC energy. \cite{SOC_1,SOC_2,SOC_agreement_1,SOC_agreement_2} \textcolor{black}{We consider LDA because it is computationally less expensive and therefore allows us to perform DFT calculations on large supercells. A drawback of the generalized gradient approximation (GGA) is that the Atomistix Toolkit 2015.1 gives rise to an indirect band gap for SL TMDCs, which is in contradiction to the already established results for TMDCs. Nonetheless, we obtain approximately the same values for both band gap and SOC using either LDA or GGA.} Figures~\ref{fig:S_S2_vacancy} and \ref{fig:M_vacancy} show the band structure of various SL TMDCs in the presence of vacancies. Black lines denote regular electronic states within the valence or conduction bands while colored lines denote the LDS. Vertical arrows show some of the allowed optical transitions observed in the optical spectra (see Fig.~\ref{fig:OS_X_X2_M-vacancies}).

\begin{table}[tp]%
\caption{Calculated in-plane and out-of-plane band gaps ${E_{g}}_{\parallel}$ and ${E_{g}}_{\perp}$ and spin-orbit splitting $\Delta_{SO}$ of the highest occupied valence band at K point.}
\centering%
\begin{tabular}{|c| c| c| c| c|}\hline
                                                      System     &    MoS$_{2}$    &    WS$_{2}$    &     MoSe$_{2}$    &     WSe$_{2}$\\ 
\hline
 ${E_{g}}_{\parallel}$[eV]                      & 1.716        & 1.684                      & 1.438                & 1.37             \\
\hline
 ${E_{g}}_{\perp}$[eV]           & 3.109        & 3.263                      & 2.516          & 2.66              \\
\hline
 $\Delta_{SO}$[meV]                      & 150        & 438                     & 195           & 482                 \\
\hline
\end{tabular}
\label{Pristine_table} 
\end{table}

\section{Tight Binding Model (TBM) and Symmetries}
\label{sec:TBM}

\subsection{General considerations}
The simplest qualitative model that can explain the existence of LDS in the band structure due to VDs is the TBM. Within the TBM aproximation the electron wavefunction can be presented as 
 $\ket{\psi} = \sum_{j,\mu \in O_j} \psi_\mu^{(j)} \varphi_\mu^{(j)}(\rr - \mathbf{R}^{(j)})$,
where $j$ enumerates atomic positions surrounding the vacancy and $\mu$ runs over the atomic orbitals $O_{j}$. In our tight binding analysis only the atoms surrounding the VD are considered in order to make the calculations simple enough for capturing the essential physical properties of the problem. The three VDs can be classified into two groups on the basis of symmetries. The $X$-vacancy lacks spatial inversion symmetry with respect to the $M$-plane of atoms, i.e. the $\sigma_{h}$ symmetry is broken, and is therefore described by the group $C_{3v}$. In contrast, the $X_{2}$ and $M$-vacancy preserve the $\sigma_{h}$ symmetry of the crystal and thus can be described by the group $D_{3h}$.\textcolor{black}{\cite{Cheiwchammangij_SOC,Song_SOI}} For the latter the electronic states break down into even and odd parity with respect to the $\sigma_{h}:z \mapsto -z$ symmetry. $d$-orbitals of the transition metal and $p^{(t,b)}$- orbitals ($t$ and $b$ denoting the top and bottom layers) of the chalcogen atoms  give the largest contribution to the conduction and valence band structure of TMDCs. \textcolor{black}{\cite{SOC_agreement_1,Guinea_tight_binding_model}} Based on the $\sigma_{h}$ symmetry, the even and odd atomic orbitals are spanned by the bases $\{d_{x^2 - y^2}, d_{xy}, d_{z^2},{~}p_{x,y}^{e}=(p_{x,y}^{(t)} + p_{x,y}^{(b)})/\sqrt{2}, {~}p_{z}^{e}=(p_{z}^{(t)} - p_{z}^{(b)})/\sqrt{2} \}$  and $\{d_{xz}, d_{yz}, {~}p_{x,y}^{o}=(p_{x,y}^{(t)} - p_{x,y}^{(b)})/\sqrt{2}, {~}p_{z}^{o}=(p_{z}^{(t)} + p_{z}^{(b)})/\sqrt{2} \}$, respectively. We also include the effects of intrinsic SOC of the form $\sim \textbf{L}\cdot\textbf{S}$. The resulting spin-orbit states transform according to irreducible representations (IRs) of the double groups $C^D_{3v}$ and $D^D_{3h}$. Group representation theory is an efficient tool for determining the allowed optical transitions across the Fermi level in solids. This will be discussed in detail in the last section. The aim of this section is to present a qualitative description of LDS appearing in the band structure (Fig:~\ref{fig:BS_pristine} b, c, d). Here following Refs.~\onlinecite{SOC_2,Guinea_tight_binding_model,First_paper}  we first develop the TBM Hamiltonian by allowing the hopping between atomic orbitals of the atoms surrounding the VD only. Also we consider a large supercell in order to suppress the intervacancy couplings. Consequently, the effects of SOC are considered as VD onsite couplings.  

\begin{figure}
\centering
\includegraphics[width=8cm]{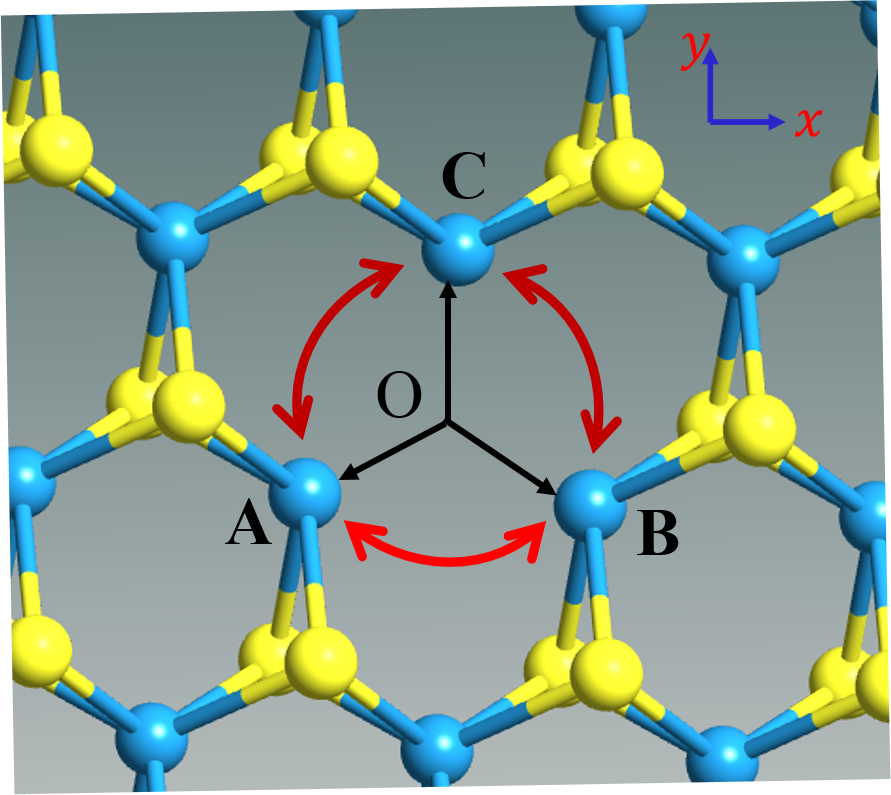}
\caption{$X_{2}$-vacancy surrounded by $M$ atoms at atomic sites $A, B$ and $C$. The defect has a rotational symmetry at angles $2\pi/3$ (or $4\pi/3$), along $z$-axis passing through $O$.}
\label{fig:X2_vacancy}
\end{figure}

\subsection{X- and X$_{2}$-vacancy}
 Both X- and X$_{2}$-vacancies are surrounded by 3 M atoms, as shown in Fig.~\ref{fig:X2_vacancy}. As mentioned earlier for M atoms, $d$-orbitals provide the main contribution. Considering 3 atomic sites $A, B, C$ with 5 $d$-orbitals on each site, we have 15 species of $d$-electrons. We will suppress the spin indices and denote electron operators collectively as a vector by $\psi=(\psi_{1}, \psi_{2}, \psi_{3}, \psi_{4}, \psi_{5})$, with $\psi_{\tau}=(d_{\tau}^A, d_{\tau}^B, d_{\tau}^C)$, where $d_{\tau}^P$ denotes the annihilation operator of electrons for orbital $\tau$ at site $P$ with $\tau=1, 2, 3, 4, 5$ standing for $d_{z^2}, d_{xy}, d_{x^2-y^2}, d_{xz}, d_{yz}$, respectively. The spinless representation of the Hamiltonian can be expressed in block form as
\begin{equation}\label{eq_TBM}
\hat{H}^{TBM}_{X_2} = 
	 \left(
	 \begin{array}{cc}
	 \hat{H}_{e}^{X_2} & \hat{0}_{9 \times6} \\
	 \hat{0}_{6 \times9} & \hat{H}_{o}^{X_2}
	 \end{array}
	 \right),
\end{equation}
where $\hat{H}_{e}^{X_{2}}$ and $\hat{H}_{o}^{X_{2}}$ are $9 \times9$ and $6 \times6$ blocks with even ($e$) and odd ($o$) parity, respectively, with respect to $\sigma_{h}$, and $\hat{0}_{m \times n}$ denotes a zero matrix of dimension $m\times n$. Here we take advantage of the inversion symmetry $\sigma_{h}$ by decoupling the orbitals with opposite parities. 
\begin{figure*}[bt]
	\begin{center}
		\includegraphics[width=7in]{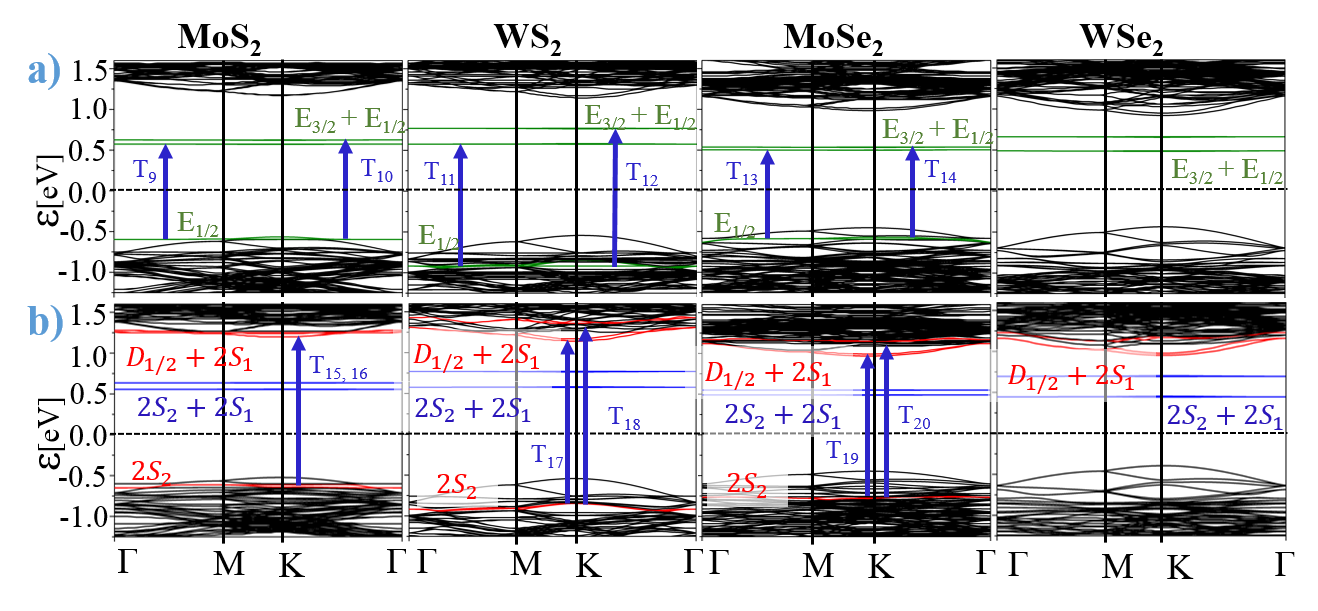}
	\end{center}
	\caption{Band structures of $7 \times 7 \times 1$ MX$_{2}$ SL TMDCs with X-vacancy (a) and X$_{2}$-vacancy (b). The Fermi level is set at $\epsilon_{F}=0$ eV. Red (blue) horizontal lines show odd (even) states w. r. to $\sigma_{h}$, while green horizontal lines in (a) represents the states with no definite symmetry. \textcolor{black}{Different LDS transform according to IRs $D_{1/2}$, $2S_{2}$ and $2S_{1}$ ($E_{1/2}$ and $E_{3/2}$) of the double group $D_{3h}^D$ ($C_{3v}^D$)}.}
	\label{fig:S_S2_vacancy}
\end{figure*}
\textcolor{black}{Also orthogonality between different orbitals on the same atomic site is enforced. The submatrices in Eq.~(\ref{eq_TBM}) are given by 
\begin{equation}\label{eq_TBM1}
\hat{H}_{e}^{X_2} = \left(
		\begin{array}{ccc}
		 \hat{H}_{e}^{1,1} & \hat{H}_{e}^{1,2} & \hat{H}_{e}^{1,3} \\
		 \hat{H}_{e}^{2,1} & \hat{H}_{e}^{2,2} & \hat{H}_{e}^{2,3}\\
		 \hat{H}_{e}^{3,1} & \hat{H}_{e}^{3,2} & \hat{H}_{e}^{3,3}
		\end{array}
\right), 
\quad
\hat{H}_{o}^{X_2} = \left(
		\begin{array}{cc}
			\hat{H}_{o}^{4,4} & \hat{H}_{o}^{4,5} \\
			\hat{H}_{o}^{5,4} & \hat{H}_{o}^{5,5}
		\end{array}
\right),
\end{equation}
where each of $\hat{H}_{e(o)}^{i,j}$ is a $3 \times 3$ matrix $i,j= {1,2,3,4,5}$. The blocks in the above Hamiltonians read
\begin{equation}\label{eq_TBM3}
\hat{H}_{e(o)}^{\alpha,\beta} = \left(
		\begin{array}{ccc}
		 \epsilon_{e(o)}^{\alpha,\beta} & t_{e(o)}^{\alpha,\beta}e^{i\theta} & t_{e(o)}^{\alpha,\beta}e^{-i\theta}\\
		 t_{e(o)}^{\alpha,\beta}e^{-i\theta} & \epsilon_{e(o)}^{\alpha,\beta} & t_{e(o)}^{\alpha,\beta}e^{i\theta}\\
		 t_{e(o)}^{\alpha,\beta}e^{i\theta} & t_{e(o)}^{\alpha,\beta}e^{-i\theta} & \epsilon_{e(o)}^{\alpha,\beta}
		\end{array}
		\right).
\end{equation} 
The diagonal elements ${\epsilon_{e(o)}^{\alpha,\beta}}$ (0 for $\alpha\neq\beta$) and the off-diagonal elements $t_{e(o)}^{\alpha,\beta}$ are phenomenological parameters describing the onsite energy and hopping between $d$-orbitals at different atomic sites, respectively}. It can be easily shown that Hamiltonian (\ref{eq_TBM3}) is invariant under $\sigma_{h}$ and  C$_{3}$ symmetry operations, for $\theta=0,\pm 2\pi/3$. But,  in addition to $\sigma_{h}$ and C$_{3}$ symmetry operations, D$_{3h}$ group also contains $\sigma_{v}$ symmetry operations, i.e. reflection by the planes perpendicular to the xy-plane and passing through the lines OA, OB and OC (Fig.~\ref{fig:X2_vacancy}). $\sigma_{v}$ demands all the complex factors appearing in Equation~(\ref{eq_TBM3}) to be 1 or equivalently $\theta=0$. Equations~(\ref{eq_TBM1}, \ref{eq_TBM3}) provides an initial insight into the nature of LDS. \textcolor{black}{One can easily show that e.g. $\hat{H}_{o}^{X_{2}}$ has a pair of 3 eigenvalues, i.e. $\bar{\epsilon}+t-\sqrt{(\delta\epsilon+h)^2+4t_{h}^{2}}$, $\bar{\epsilon}-t/2-\sqrt{(\delta\epsilon-h/2)^2+t_{h}^{2}}$, $\bar{\epsilon}-t/2-\sqrt{(\delta\epsilon-h/2)^2+t_{h}^{2}}$ and  $\bar{\epsilon}+t+\sqrt{(\delta\epsilon+h)^2+4t_{h}^{2}}$, $\bar{\epsilon}-t/2+\sqrt{(\delta\epsilon-h/2)^2+t_{h}^{2}}$, $\bar{\epsilon}+t/2-\sqrt{(\delta\epsilon-h/2)^2+t_{h}^{2}}$, where $\bar{\epsilon}$, $\delta\epsilon$ are related to addition and subtraction of onsite energies for orbitals d$_{xz}$, d$_{yz}$; $t$, $h$ are related to addition and subtraction of hopping parameters of the same orbitals at different sites, and $t_{h}$ is the hopping parameter of different orbitals at different atomic sites. Each pair contains a two-fold doublet, which explains the existence of triplets within the band structure.\cite{First_paper}} However, the apparent two-fold degeneracy, which arises from the overlap of neighboring atomic orbitals, is lifted in the presence of SOC. Here we emphasize that each $d$-orbital appears in the form of triplets in the band structure. Thus, there is a total of 15 LDS (in the absence of SOC) for the case of the X$_2$-vacancy. It may appear that the simplest TBM may contradict the numerical results in Fig.~\ref{fig:S_S2_vacancy}, where calculations show a lower number of LDS. A closer inspection of the numerical results, however, resolves this contradiction in favor of the TBM. In fact, in addition to the LDS appearing within the band gap region, there are also LDS deep inside the valence bands, with possibility to mix with the extended states in the bulk.

SOC in the Hamiltonian is included by a pure atomic term\textcolor{black}{\cite{SOC_2}} and for simplicity we consider only the onsite contribution arising from the M atoms surrounding the vacancy. Using the basis $\ket{d_{z^2},\uparrow}, \ket{d_{xy},\uparrow}, \ket{d_{x^2-y^2},\uparrow}, \ket{d_{xz},\uparrow}, \ket{d_{yz},\uparrow}$ and $\ket{d_{z^2},\downarrow}, \ket{d_{xy},\downarrow}, \ket{d_{x^2-y^2},\downarrow}, \ket{d_{xz},\downarrow}, \ket{d_{yz},\downarrow}$, we can write the SOC Hamiltonian as 
\begin{equation}\label{eq_SOI_H}
\hat{H}^{SOC}_{X_2}=\frac{\Delta}{2}\textbf{L} \cdot \textbf{S}=\frac{\Delta}{2}
											 \left(
	 										 \begin{array}{cc}
											 \hat{L}_{z} & \hat{L}_{-} \\
											 \hat{L}_{+} & -\hat{L}_{z}
											 \end{array}
											 \right),
\end{equation}
where
\begin{equation}
\hat{L}_{z} = \left(
		\begin{array}{ccccc}
		\hat{0}_{3} & \hat{0}_{3} & \hat{0}_{3} & \hat{0}_{3} & \hat{0}_{3}  \\
		\hat{0}_{3} & \hat{0}_{3} & 2i\times \hat{I}_{3} & \hat{0}_{3} & \hat{0}_{3}  \\
		\hat{0}_{3} & -2i\times \hat{I}_{3} & \hat{0}_{3} & \hat{0}_{3} & \hat{0}_{3}  \\
		\hat{0}_{3} & \hat{0}_{3} & \hat{0}_{3} & \hat{0}_{3} & -i\times\hat{I}_{3}  \\
		\hat{0}_{3} & \hat{0}_{3} & \hat{0}_{3} & i\times\hat{I}_{3} & \hat{0}_{3}  \\
		\end{array}
	 \right),
\end{equation}

\begin{equation}
\hat{L}_{+} = \left(
		\begin{array}{ccccc}
		\hat{0}_{3} & \hat{0}_{3} & \hat{0}_{3} & \sqrt{3}\times\hat{I}_{3} & i\sqrt{3}\hat{I}_{3}  \\
		\hat{0}_{3} & \hat{0}_{3} & \hat{0}_{3} & -i\times\hat{I}_{3} & -1\times\hat{I}_{3}  \\
		\hat{0}_{3} & \hat{0}_{3} & \hat{0}_{3} & -1\times\hat{I}_{3} & i\times\hat{I}_{3}  \\
		-\sqrt{3}\times\hat{I}_{3} & i\times\hat{I}_{3} & \hat{I}_{3} & \hat{0}_{3} & \hat{0}_{3}  \\
		-i\sqrt{3}\times\hat{I}_{3} & \hat{I}_{3} & -i\times\hat{I}_{3} &\hat{0}_{3} & \hat{0}_{3}  \\
		\end{array}
	 \right)
\end{equation}
and $\hat{L}_{-}=\hat{L}_{+}^{\dagger}$. The off-diagonal elements $\hat{L}_{\pm}$ in Eq.~\eqref{eq_SOI_H} couple the even to the odd blocks of the Hamiltoian matrix shown in Eq.~\eqref{eq_TBM} and are related to the spin flip processes due to the SOC, which give rise to virtual transitions.\cite{Guinea_SOI_TMDCs} Because of the large spatial anisotropy of an atomically thin layer of TMDC, for the pristine case these off-diagonal terms can be neglected, which is substantiated by our DFT calculations (see below). A generalized SOC states has the form 
\begin{equation}\label{Generalized_SOC}
\ket{\Psi} = \alpha\ket{\zeta}\ket{\uparrow}+\beta\ket{\xi}\ket{\downarrow}. 
\end{equation}
Here, $\ket{\zeta}$ and $\ket{\xi}$ are orbital states of the spin-up and spin-down states $\ket{\uparrow}$ and $\ket{\downarrow}$, respectively, and $\alpha$, $\beta$ are probablility amplitudes for the up and down spinors.  DFT calculations reveal that for SOC  Bloch states corresponding to LDS, either $\alpha\ll\beta$ or $\alpha\gg\beta$ in the majority of cases (Fig:~\ref{fig:Bloch_state_S_vacancy}), corresponding to strong polarizations of the LDS in z direction (b), which is due to the large spatial anisotropy. It can be calculated that LDS are spin polarized for X$_{2}$-vacancy or the Bloch states for X$_{2}$-vacancy preserve the $\sigma_{h}$ symmetry. Therefore we anticipate that the effects of $\hat{L}_{\pm}$ can safely be neglected for the X$_{2}$-vacancy. The full tight binding Hamiltonian can be written as 
\begin{equation}
\begin{split}
\hat{H}_{X_2} & =I_{2}\otimes \hat{H}_{X_{2}}^{TBM}+\hat{H}_{X_{2}}^{SOC} \\
                       & =
	 \left(
	 \begin{array}{cc}
	 \hat{H}_{X_{2}}^{TBM}+ \frac{\Delta}{2}\hat{L}_{z} & \hat{0}_{9 \times9} \\
	 \hat{0}_{9 \times9} & \hat{H}_{X_{2}}^{TBM}- \frac{\Delta}{2}\hat{L}_{z}
	 \end{array}
	 \right).
\end{split}
\end{equation}
The Hamiltonian appears to be block diagonal, which indicates that spin states in $z$ direction are not mixed by spin flip processes and therefore the spin in $z$ direction is still a good quantum number due to $\sigma_{h}$ symmetry. As mentioned above, the abscence of spin flip processes can be attributed to the 2D character of TMDCs or due to the large anisotropy between $xy$-plane and $z$-axis. In case of the $X$-vacancy, due to lack of $\sigma_{h}$ symmetry, defect states appear with no definite parity (Fig:~\ref{fig:Bloch_state_S_vacancy}) (a). Therefore, here we argue that for the $X$-vacancy the off-diagonal terms $\hat{L}_{\pm}$ in Eq. \eqref{eq_SOI_H} need to be taken into account. 

In the absence of SOC each energy band is doubly degenerate (spin-up and spin-down states at each k point). SOC lowers the symmetry and can break the spin degeneracy at k points away from high symmetry points. However, time reversal symmetry leads to the condition that $\varepsilon(\textbf{k},\uparrow)=\varepsilon(-\textbf{k},\downarrow)$, commonly known as Kramers degeneracy. This degeneracy is reflected in the band structure (Fig.~\ref{fig:S_S2_vacancy}), where each energy level is doubly degenerate for both types of vacancies. In solids or 2D surfaces spin splitting depends both on the size of atomic SOC and of the gradient of electric potential.\cite{Petersen200049} This difference in the gradient of electric potential leads to the different spin splittings for same types of defects in different TMDCs as shown in Fig.~\ref{fig:S_S2_vacancy} and in  Table~\ref{table_SOC_Splitting}.

\begin{figure*}[bt]
	\begin{center}
		\includegraphics[width=7in]{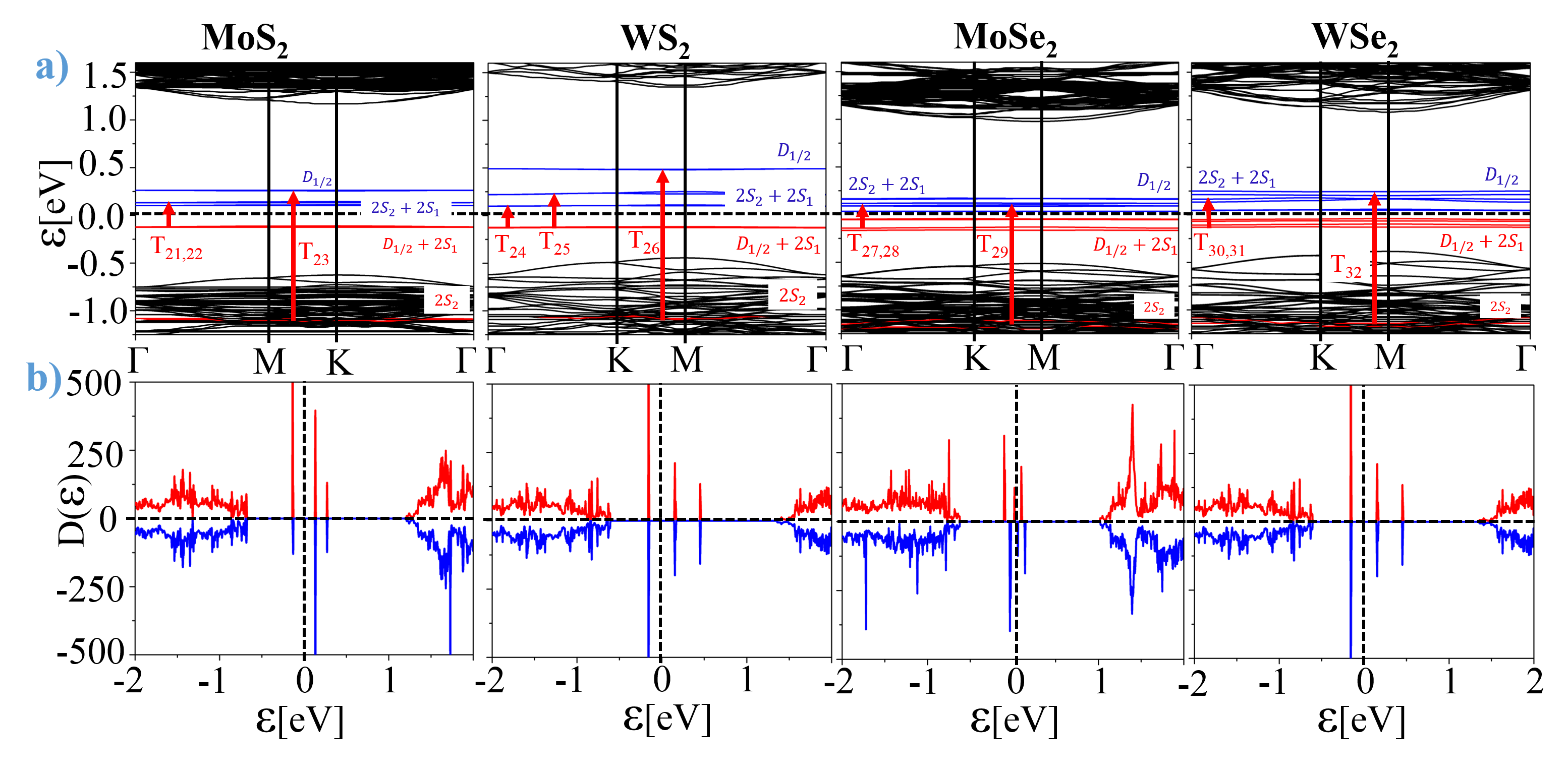}
	\end{center}
	\caption{(a)Band structures of $7 \times 7 \times 1$ MX$_{2}$ SL TMDCs with M-vacancy. The Fermi level is set at $\epsilon_{F}=0$ eV. Red (blue) lines show odd (even) states w. r. to $\sigma_{h}$. \textcolor{black}{Different LDS transform according to IRs $D_{1/2}$, $2S_{2}$ and $2S_{1}$ of double group $D_{3h}^D$} (b) Spin polarized density of states of $7 \times 7 \times 1$ MX$_{2}$ SL TMDCs with M-vacancy for spin up(down) are shown in red(blue).}
	\label{fig:M_vacancy}
\end{figure*}

\subsection{M-vacancy}
There are 6 chalcogen X atoms in the top and bottom layers, surrounding the tranisition metal M-vacancy. Thus, there are 18 species of electrons corresponding to six possible combinations of $p$-orbitals (3 of them even and 3 odd with respect to $\sigma_{h}$) at 3 in-plane atomic positions. Proceeding as before the TBM for the M-vacancy can be written as
\textcolor{black}{
\begin{eqnarray}
\hat{H}^{TBM}_{M} & = & 
	 \left(
	 \begin{array}{cc}
	 \hat{H}_{e}^{M} & \hat{0}_{9 \times9} \\
	 \hat{0}_{9 \times9} & \hat{H}_{o}^{M}
	 \end{array}
	 \right),
\label{eq_TBM_M}\\
\hat{H}_{e}^{M} & = & \left(
		\begin{array}{ccc}
		 \hat{H}_{e}^{6,6} & \hat{H}_{e}^{6,7} & \hat{H}_{e}^{6,8} \\
		 \hat{H}_{e}^{7,6} & \hat{H}_{e}^{7,7} & \hat{H}_{e}^{7,8}\\
		 \hat{H}_{e}^{8,6} & \hat{H}_{e}^{8,7} & \hat{H}_{e}^{8,8}
		\end{array}
\right), \label{eq_TBM_M2}\\
\hat{H}_{o}^{M} & = &  \left(
		\begin{array}{ccc}
		 \hat{H}_{o}^{9,9} & \hat{H}_{o}^{9,10} & \hat{H}_{o}^{9,11} \\
		 \hat{H}_{o}^{10,9} & \hat{H}_{o}^{10,10} & \hat{H}_{o}^{10,11}\\
		 \hat{H}_{o}^{11,9} & \hat{H}_{o}^{11,10} & \hat{H}_{o}^{11,11}
		\end{array}
\right), \label{eq_TBM_M3}
\end{eqnarray}
} 
where each of $\hat{H}_{e(o)}^{j}$ is a $3 \times 3$ matrix corresponding to even (odd) combinations of $p$ orbitals, with $j= {6,7,8,9,10,11}$ being indices reserved for the ${p_{x}^{e},p_{y}^{e},p_{z}^{e},p_{x}^{o},p_{y}^{o},p_{z}^{o}}$ orbitals, respectively. Each $\hat{H}_{e(o)}^{\beta}$ in Eq.~(\ref{eq_TBM_M2}) has the same form as in Eq.~(\ref{eq_TBM3}).  

\begin{figure*}[bt]
	\begin{center}
		\includegraphics[width=6.5in]{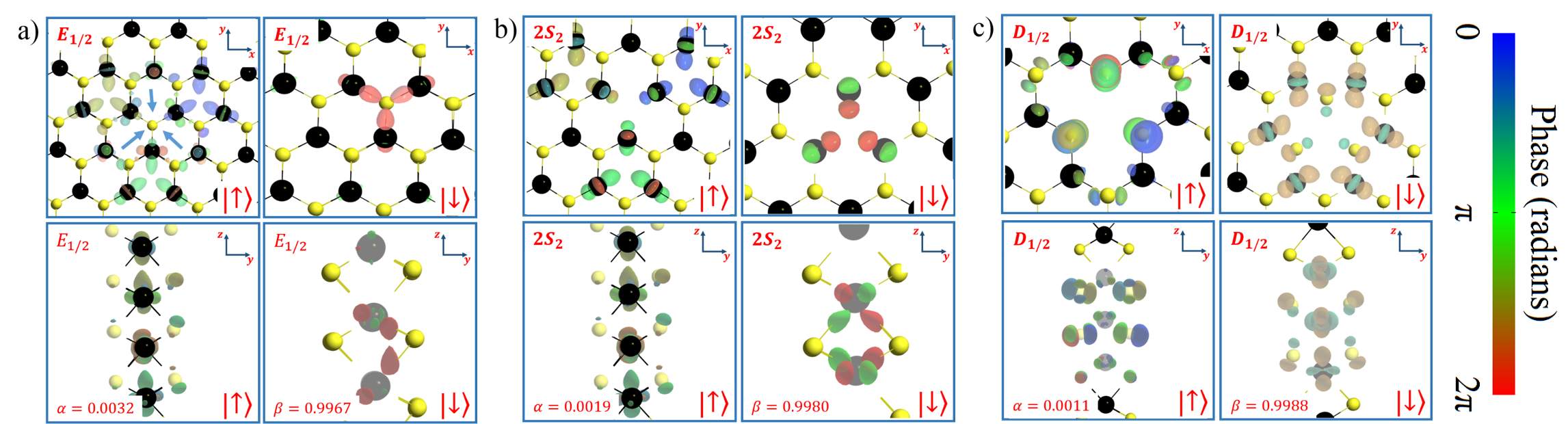}
	\end{center}
	\caption{Bloch states corresponding to a LDS a) X and b) X$_{2}$-vacancy c) M-vacancy in MoS$_{2}$. $E_{1/2}$ and $2S_{2}$ and D$_{1/2}$are corresponding IRs. Both the top ($xy$-plane) and side ($yz$-plane) views are presented. Bloch states are spin polarized in each case $\beta \gg\alpha$.}
	\label{fig:Bloch_state_S_vacancy}
\end{figure*}

\begin{figure*}[bt]
	\begin{center}
		\includegraphics[width=6.5in]{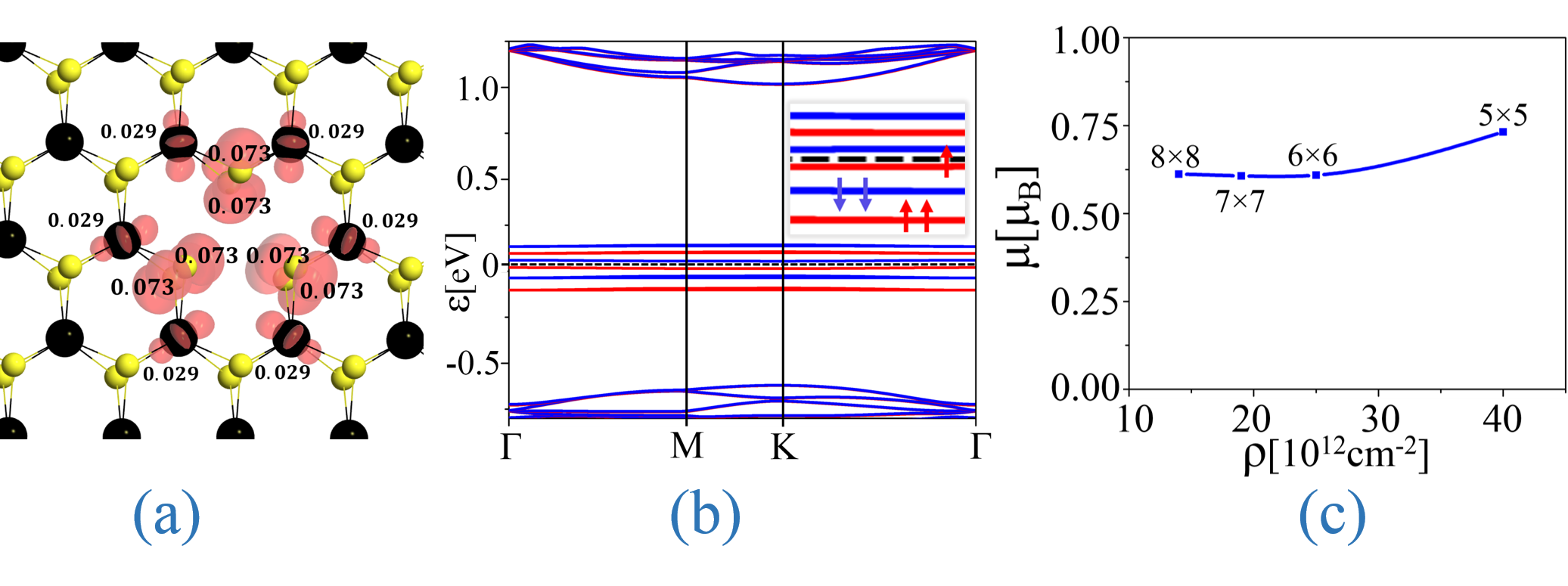}
	\end{center}
	\caption{(a) Isosurface plot of the spin density of MoSe$_{2}$ monolayer with one Mo vacancy. The black and yellow balls represent Mo and Se atoms, respectively. The red isosurface shoes the spin density. Magnetic moment $\mu$ at neighbouring and next neighbouring sites surrounding the vacancy  is also calculated which gives a total $\mu$ of 0.6 in units of $\mu_{B}$. (b) Spin polarized band structure of $7\times 7$ MoSe$_{2}$ with one Mo vacancy, red and blue lines showes states for spin up and down, respectively. (c) Magnetic moment $\mu$ vs defect density $\rho$.}
	\label{fig:SD_MP_SPBS_MD}
\end{figure*}

SOC is included by considering the Hamiltonian described in Eq.~(\ref{eq_SOI_H}). The M-vacancy also preserves the $\sigma_{h}$ symmetry. Again, the $\hat{L}_{z}S_z$ term gives the largest contribution due to the large anisotropy. The Bloch states are shown in Fig.~\ref{fig:Bloch_state_S_vacancy}, from which it can be concluded that LDS are spin polarized also in the case of M-vacancy. The matrices for operators $\hat{L}_{z}$ and $\hat{L}_{\pm}$ in the case of the M-vacancy can be written as
\begin{equation}
\hat{L}_{z} = \left(
		\begin{array}{cccccc}
		\hat{0}_{3} & -i\times \hat{I}_{3} & \hat{0}_{3} & \hat{0}_{3} & \hat{0}_{3} & \hat{0}_{3}  \\
		i\times \hat{I}_{3} & \hat{0}_{3} & \hat{0}_{3} & \hat{0}_{3} & \hat{0}_{3} & \hat{0}_{3}  \\
		\hat{0}_{3} & \hat{0}_{3} & \hat{0}_{3} & \hat{0}_{3} & \hat{0}_{3} & \hat{0}_{3}  \\
		\hat{0}_{3} & \hat{0}_{3} & \hat{0}_{3} & \hat{0}_{3} & -i\times\hat{I}_{3} & \hat{0}_{3}  \\
		\hat{0}_{3} & \hat{0}_{3} & \hat{0}_{3} & i\times\hat{I}_{3} & \hat{0}_{3} & \hat{0}_{3}  \\
		\hat{0}_{3} & \hat{0}_{3} & \hat{0}_{3} & \hat{0}_{3} & \hat{0}_{3} & \hat{0}_{3}  \\
		\end{array}
	 \right),
\end{equation} 
,
\begin{equation}
\hat{L}_{+} = \left(
		\begin{array}{cccccc}
		\hat{0}_{3} & \hat{0}_{3} & \hat{0}_{3} & \hat{0}_{3} & \hat{0}_{3} & -1 \times\hat{I}_{3}  \\
		\hat{0}_{3} & \hat{0}_{3} & \hat{0}_{3} & \hat{0}_{3} & \hat{0}_{3} & i\times \hat{I}_{3}  \\
		\hat{0}_{3} & \hat{0}_{3} & \hat{0}_{3} & \hat{I}_{3} & -i\times \hat{I}_{3} & \hat{0}_{3}  \\
		\hat{0}_{3} & \hat{0}_{3} & -1\times\hat{I}_{3} & \hat{0}_{3} & \hat{0}_{3} & \hat{0}_{3}  \\
		\hat{0}_{3} & \hat{0}_{3} & i\times\hat{I}_{3} & \hat{0}_{3} & \hat{0}_{3} & \hat{0}_{3}  \\
		 \hat{I}_{3} &- i\times \hat{I}_{3} & \hat{0}_{3} & \hat{0}_{3} & \hat{0}_{3} & \hat{0}_{3}  \\
		\end{array}
	 \right),
\end{equation} 
and $\hat{L}_{-}=\hat{L}_{+}^{\dagger}$. The DFT calculations show that the Kramers degeneracy is preserved for MoS$_{2}$ and WS$_{2}$ while it is broken for MoSe$_{2}$ and WSe$_{2}$. In Ref.~\onlinecite{magnetic_Mo_MoSe2}, it has been shown that presence of Mo vacancies in MoSe$_{2}$ can induce spin polarization and results in long range antiferromagnetic coupling between local magnetic moments, even at a distance above {13 \AA} due to the large spatial extensions of spin density. \textcolor{black}{The local magnetic moment on each M-vacancy breaks the time reversal symmetry and therefore lifts Kramers degeneracy in MoSe$_{2}$ in the presence of Mo vacancies. In Fig.~\ref{fig:M_vacancy} (b) we show plots for the density of states (DOS) obtained by the local density spin approximation (LSDA) method for the M-vacancy in different TMDCs, in order to confirm that indeed the Mo-vacancy in MoSe$_{2}$ exhibits magnetic signatures. Our spin-polarized DFT calculations show that the exchange correlation potential leads to a spin splitting only for the MoSe$_{2}$ system. In Fig.~\ref{fig:SD_MP_SPBS_MD}(a) the isosurface plot for the spin density is shown along with the magnetic moment $\mu$ calculated by means of the Mulliken population analysis at all the nearest-neighboring atomic sites (Se atoms) and the next-nearest neighboring sites (Mo atoms) surrounding the vacancy. Our results show that the main contribution to the magnetism is due to the $p$-orbitals localized at the Se atoms and the $d$-orbitals localized at the next-nearest Mo atoms surrounding the Mo vacancy. The electronic band structure calculated by using LSDA is shown in  Fig.~\ref{fig:SD_MP_SPBS_MD}(b). The spin states of the LDS are split whereas the bulk states do not show any magnetic moment. The magnetic moment of the LDS is governed by the unpaired electron spins according to Hund's rules, which is 1 $\mu_{B}$ (inset of Fig.~\ref{fig:SD_MP_SPBS_MD}(b)). The calculations yield a slightly smaller magnetic moment of 0.6$\mu_{B}$, which is acceptable within standard DFT limits. Finally, we plot the magnetic moment $\mu$ vs defect density $\rho$ in Fig.~\ref{fig:SD_MP_SPBS_MD}(c), where the localization of the magnetic moment is demonstrated for densities $\rho<25\times 10^{12}$ cm$^{-2}$}. We also notice splittings in the case of WSe$_{2}$, but no magnetic signatures are found. We attribute this splitting as a result of interaction between adjacent vacancies due to large spatial extensions of W and Se orbitals. 

\begin{table}[tp]%
\caption{SOC splitting $\Delta_{X, X_{2}, M}$ in LDS appearing above the Fermi level for X, X$_{2}$ and M vacancies respectively, in different TMDCs}
\centering%
\begin{tabular}{|c| c| c| c| c|}\hline
                                                  System         &    MoS$_{2}$    &    WS$_{2}$    &     MoSe$_{2}$    &     WSe$_{2}$\\ 
\hline
 $\Delta_{X}$[meV]                      & 51        & 192                      & 34                & 173             \\
\hline
 $\Delta_{X_{2}}$[meV]           & 78        & 278                      & 60          & 251              \\
\hline
 $\Delta_{M}$[meV]                      & 32.5        & 130                     & 54           & 100                 \\
\hline
\end{tabular}
\label{table_SOC_Splitting} 
\end{table}

\section{Defect states within the Dirac equation formalism}
\label{sec:Dirac}
The main analytical tool for describing properties of electron states in transition metals monolayers is the Dirac equation, which emerges within the $k\cdot p$-approximation as the two-band model.\textcolor{blue}{\cite{SOC_1,Kormanyos_k_P,Kormanyos_defects}} Such description is valid when the main role is played by low lying excitations near the band edges. This assumption, however, is ill-justified for the case of vacancies and, indeed, as will be demonstrated below the Dirac equation fails to reproduce many important features of the defect states. At the same time, the Dirac equation allows one to establish several important features and, first of all, the sole existence of the states bound to vacancies. For example, within the framework provided by the Schrodinger equation the vacancy is naturally represented by a repulsive potential, which cannot support bound states. The Dirac equation, in turn, provides more options for describing defects and, as will be shown below, when the special boundary conditions are enforced at the boundary of the defect, the states localized near the boundary appear.

The formalism of Dirac equation can be introduced as follows. Within the 
$k\cdot p$-approximation the electron states are described by $\Psi(\rr) e^{i \mathbf{K}\cdot\rr}$, where $\Psi(\rr)$ is a smooth function of coordinates. Adopting the two-band approximation, $\Psi(\rr)$ is presented as a two-component spinor, which satisfies a 2D Dirac-like equation. For example, for MoS${}_2$ the two-band approximation is often implemented retaining only the dominating contribution of Mo's $d$ orbitals,\cite{SOC_1, SOC_2} so that near the inequivalent $K$-points of the Brillouin zone, $\mathbf{K}_\tau = \tau \mathbf{K}$ with $\tau = \pm 1$, spinors $\Psi_\tau(\rr)$ are spanned by $\ket{d_{x^2 - y^2} - i \tau d_{xy}}$ and $\ket{d_{z^2}}$, the states representing the top of the valence band and the bottom of the conduction band, respectively. Spinors $\Psi_\tau$, thereby, satisfy $\left[\sigma_z \Delta + v \left(\tau \sigma_x p_x + \sigma_y p_y\right)\right]\Psi_\tau(\rr) = \epsilon \Psi(\rr)$, where $2\Delta$ is the width of the gap and $\epsilon$ is the energy counted from the center of the gap $\epsilon_c$. 

In order to eliminate the valley dependence of the Hamiltonian governing the spatial distribution of $\Psi_\tau$, it is convenient to introduce $\Phi_+ = \Psi_+$ and $\Phi_- = \sigma_y\Psi_-$, which satisfy $\HH_\tau \Phi_\tau = \epsilon \Phi_\tau$, where
\begin{equation}\label{eq:Ham_xi}
\HH_\tau = \tau \sigma_z \Delta + v \bosi \cdot \mathbf{p}.
\end{equation}
Thus solutions for electrons in different valleys are related by simple reverting the sign of $\Delta$. Combining $\Phi_{\pm}$ into a single 4-spinor
\begin{equation}\label{eq:4sp_def}
 \Phi = \Phi_+ \oplus \Phi_-
\end{equation}
the equations of motion for different valleys can be presented in a unified form $\HH \Phi = \epsilon \Phi$ with
\begin{equation}\label{eq:4ham}
 \HH = \tau_z \otimes \sigma_z  \Delta + v \tau_0 \otimes \bosi \cdot \mathbf{p},
\end{equation}
where $\tau_i$ with $i=x,y,z$ and $\tau_0$ are the Pauli matrices and the identity matrix, respectively, acting on the valley space.

Hamiltonian \eqref{eq:4ham} possesses the cylindrical symmetry, which can be employed by presenting $\bosi \cdot \mathbf{p} = - i \sigma_r \partial/\partial r - i r^{-1}\sigma_\phi \partial/\partial \phi$, where
\begin{equation}
\sigma_r = \left(
		\begin{array}{cc}
		 0 & e^{-i \phi} \\
		 e^{i \phi} & 0
		\end{array}
\right), 
\quad
\sigma_\phi = \left(
		\begin{array}{cc}
			0 & -i e^{-i \phi} \\
			i e^{i \phi} & 0
		\end{array}
\right).
\end{equation}
The explicit angular dependence is eliminated by introducing $\widetilde{\Phi}_\tau = e^{i \sigma_z \phi/2} \Phi_\tau$, which accounts for rotation of the spinors $\Phi_\tau$, while encircling the origin. It should be noted that due to the relation $\Phi_- = \sigma_y \Psi_-$ the rotation directions of $\Psi_+$ and $\Psi_-$ are different: $\widetilde{\Psi}_{\tau} = e^{i \tau \sigma_z \phi/2} \Psi_\tau$. Thus the winding numbers of spinors corresponding to electrons belonging to different valleys have opposite signs.

Spinors $\widetilde{\Phi}_\tau$ satisfy
\begin{equation}\label{eq:rotated}
 \widetilde{\HH}_\tau \widetilde{\Phi}_\tau = \epsilon \widetilde{\Phi}_\tau,
\end{equation}
where $\widetilde{\HH}_\tau = e^{i \sigma_z \phi/2} \HH_\tau e^{-i \sigma_z \phi/2}$ has the form
\begin{equation}\label{eq:H_rot}
\widetilde{\HH} = \tau \sigma_z \Delta - i v \left[\sigma_x \left(\frac{\partial}{\partial r} + \frac{1}{2r}\right) + \sigma_y \frac{1}{r}\frac{\partial}{\partial \phi} \right].
\end{equation}

Equation \eqref{eq:rotated} is solved by separating variables $\widetilde{\Phi}_\tau(r, \phi) = \sum_{m = - \infty}^\infty \widetilde{\Phi}_{\tau;m}(r) e^{i m \phi}$. For amplitudes $\widetilde{\Phi}_{\tau;m}(r)$, we find the general solution
\begin{equation}\label{eq:gen_sol}
\begin{split}
 \widetilde{\Phi}_{\tau;m} & (r) =  \sqrt{Q r}  \\
 \times & \left(
 	\begin{array}{c}
 	\frac{1}{\sqrt{\epsilon - \tau \Delta}} \left(a_{\tau;m} h_{m-1}^{(1)}(Q r) + b_{\tau;m} h_{m-1}^{(2)}(Q r) \right) \\
 	\frac{i}{\sqrt{\epsilon + \tau \Delta}} \left(a_{\tau;m} h_{m}^{(1)}(Q r) + b_{\tau;m} h_{m}^{(2)}(Q r) \right)
 	\end{array}
 \right),
\end{split}
\end{equation}
where $h_{m}^{(1,2)}(Qr)$ are the spherical Hankel functions, $Q=v^{-1}\sqrt{\Delta^2 - \epsilon^2}$, $a_{\tau;m}$ and $b_{\tau;m}$ are arbitrary constants.

We are interested in bound states and, therefore, in solutions of Eq.~\eqref{eq:rotated} corresponding to energies inside the gap. For such energies, we have $Q = i \kappa$ with non-negative $\kappa = \sqrt{\Delta^2 - \epsilon^2}$. From the regularity condition at infinity, it follows that $b_\tau^{(m)} = 0$, while $a_\tau^{(m)}$ are determined from the normalization condition. The solution can be written as
\begin{equation}\label{eq:sol_gap}
 \widetilde{\Phi}_{+,m} = N_{+,m}
	 \left(
	 \begin{array}{c}
	 \frac{1}{\sqrt{\Delta - \epsilon}} g_{m-1}(\kappa r) \\
	 \frac{i}{\sqrt{\Delta + \epsilon}} g_{m}(\kappa r)
	 \end{array}
	 \right),
\end{equation}
where we have denoted the normalization constant by $N_{+,m}$. The functions $g_m(z)$ are related to the modified spherical Hankel functions $g_m(z) = 2 k_m(z)/\pi$ and for $m>0$ can be presented as
\begin{equation}\label{eq:g_expr}
 g_m(z) = (-z)^m \left(\frac{d}{zdz}\right)^m \frac{e^{-z}}{z}.
\end{equation}
Taking into account the relation
\begin{equation}\label{eq:g_sym}
g_{-m}(z) = g_{m-1}(z),
\end{equation}
we can use Eq.~\eqref{eq:g_expr} for finding $g_m(z)$ with $m<0$ as well.
With the help of this relation, one can show, starting from Eq.~\eqref{eq:gen_sol}, that
\begin{equation}\label{eq:pos_neg_phi}
 \widetilde{\Phi}_{-,-m} = c \sigma_x \widetilde{\Phi}_{+,m},
\end{equation}
where $c$ is a phase factor, $|c| = 1$. Such connection between solutions corresponding to electrons from different valleys allows us to limit our attention to $\tau = +1$.

\begin{figure*}
\centering
\includegraphics[width=14cm]{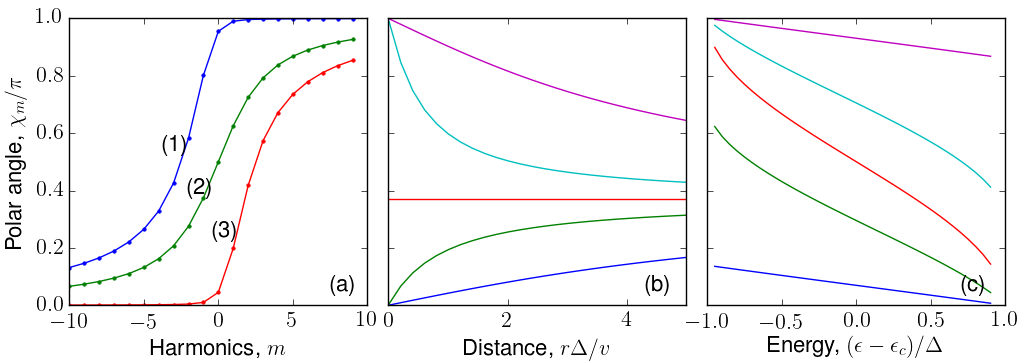}
\caption{The polar angle $\chi_m$ characterizing the defect state. (a) Dependence of $\chi_m$ on $m$. Curves (1), (2) and (3) correspond to $(\epsilon - \epsilon_c)/\Delta = -0.99, 0, 0.99$, respectively, where $\epsilon_c$ stands for the center of the gap. The distance from the center of the anti-dot is taken $r \Delta /v = 1$ (it is assumed that $r > r_0$). (b) For the fixed energy $(\epsilon - \epsilon_c)/\Delta = 0.4$ the dependence $\chi_m$ on $r$ is shown for $m$; from top to bottom $m = 5, 1, 0, -1, -5$. (c) For $r \Delta/v = 1$ the dependence of $\chi_m$ on energy is shown for the same set of $m$ as in (b).}
\label{fig:angles}
\end{figure*}

The functions $g_m(z)$ can be shown to be non-negative. Thus, we can rewrite
\begin{equation}\label{eq:Phi_sc}
\widetilde{\Phi}_{+,m} = \widetilde{N}_{+,m}
	 \left(
	 \begin{array}{c}
	 \cos(\chi_m/2) \\
	 i \sin(\chi_m/2)
	 \end{array}
	 \right),
\end{equation}
with $0 \leq \chi_m \leq \pi$. This representation shows that at any chosen distance from the center of the vacancy, the defect states have the form of a spin coherent state \cite{aravind_spin_1999,CombescureCoherent2012} lying in the plane perpendicular to $\mathbf{n}_B$, the vector normal to the boundary of the anti-dot and directed outward. The angle $\chi_m = 2 \arctan(F_m)$, where
\begin{equation}\label{eq:chi_ang}
 F_m = 
 \frac{g_m(\kappa r) \sqrt{\Delta - \epsilon}}
 {g_{m-1}(\kappa r) \sqrt{\Delta + \epsilon}},
\end{equation}
has the meaning of the polar angle of the vector characterizing the direction of the spin coherent state. Its dependence on $m$ is illustrated by Fig.~\ref{fig:angles}, which shows that $\chi_m$ monotonously increases from $\chi_{-\infty} = 0$ to $\chi_\infty = \pi$. It is also a monotonous function of $r$ (increasing for $m < 0$ and decreasing for $m > 0$) and monotonously increasing function of energy. Taking into account Eq.~\eqref{eq:g_sym} one can see the important symmetry
\begin{equation}\label{eq:F_m_flip}
 F_m(\epsilon) = 1/F_{-m}(-\epsilon).
\end{equation}

Thus, the states with $\tau = 1$ lie in the half plane corresponding to the positive projection on the vector tangent to the boundary $\mathbf{e}_\phi = \mathbf{e}_z \times \mathbf{n}_B$, while the states with $\tau = -1$ have the opposite orientation, due to $\braket{\widetilde{\Phi}_{-,m}}{\widetilde{\Phi}_{+,m} } = 0$, which can be easily checked.

The energies of the defects states (and their sole existence) are determined by the boundary condition on the boundary of the anti-dot. The general form of the condition is found requiring that the radial component of the probability current must vanish at the boundary,\cite{Boundary_condition_1,Boundary_condition_2} $\matrixel{\Phi(r_0)}{\mathbf{n}_B \cdot \mathbf{J}}{\Phi(r_0)} = 0$, where  $\mathbf{J} = v \tau_0 \otimes \bosi$. This condition is equivalent to $M \Phi = \Phi$, where $\Phi$ is the 4-spinor defined by Eq.~\eqref{eq:4sp_def} and the Hermitian matrix $M$ has the eigenvalues $\pm1$ and anticommutes with the radial component of the current operator $\lbrace \mathbf{J} \cdot \mathbf{n}_B, M \rbrace = 0$. 
Within the infinite mass model,\cite{Boundary_condition_1,Boundary_condition_2} the anti-dot is represented as a region with renormalized width of the gap $\Delta \to \Delta(1+d(r))$ with $d(r) = 0$ for $r> r_0$ and $d(r) \to \infty$ when $r< r_0$, so that in this case $M = (\boldsymbol{\tau} \cdot \mathbf{e}_z) \otimes (\bosi \cdot \mathbf{e}_\phi)$. In other words, within this model in order to have decaying electron distribution inside the anti-dot
$\widetilde{\Phi}_\tau(r)$ must be proportional to $\ket{\tau\mathbf{e}_\phi}$ as $r$ approaches $r_0$.


The condition $\widetilde{\Phi}_{+,m}(r_0) \propto \ket{\mathbf{e}_\phi}$, or $\chi_m = \pi/2$, constitutes the condition imposed on the energy of the bound state
\begin{equation}\label{eq:ener_eq}
 F_m(\epsilon, r_0) = 1.
\end{equation}
In virtue of Eq.~\eqref{eq:F_m_flip}, if for some $m$ there exists a bound solution with the energy $\epsilon$, then there is the solution corresponding to $m' = -m$ with the energy $-\epsilon$. Thus within the infinite mass model the spectrum of the defect states is symmetric with respect to the center of the gap.

For $m = 0$, Eq.~\eqref{eq:ener_eq} has the simplest form $F_0 = \sqrt{\Delta - \epsilon}/\sqrt{\Delta + \epsilon} = 1$ with the solution
\begin{equation}\label{eq:m0_eq}
  \epsilon_0 = 0.
\end{equation}
Thus the anti-dot independently of its size supports a bound state with the energy at the center of the gap.

\begin{figure}
\centering
\includegraphics[width=8cm]{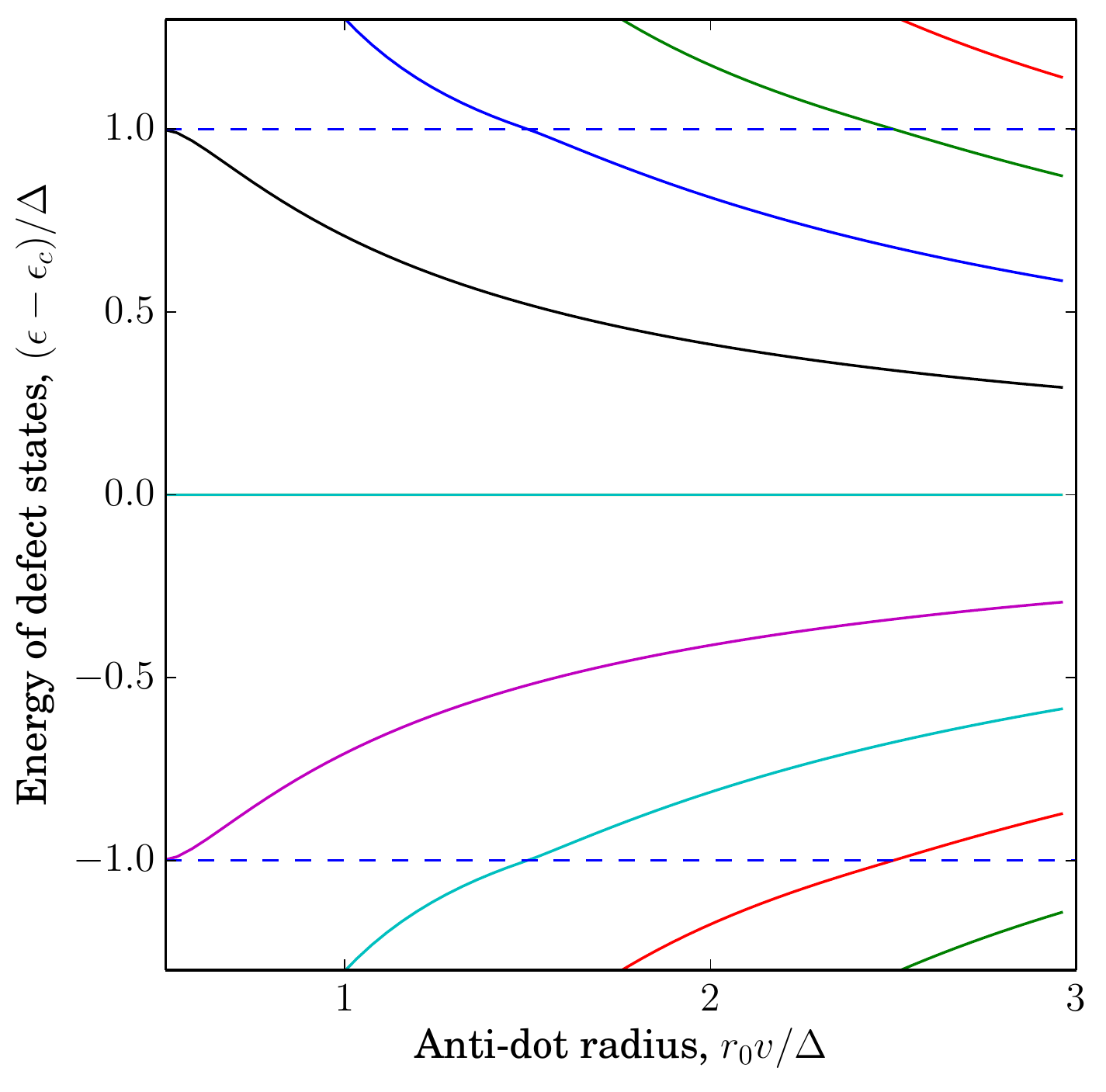}
\caption{The dependence of the spectrum of defect states on the normalized radius of the anti-dot, $r_0 v/\Delta$. The dashed lines show the edges of the gap.  The section at particular $r_0$ presents the spectrum of the defect states in the infinite mass model. The central line corresponds to $m=0$ and to states with increasing $|\epsilon - \epsilon_c|$ correspond states with increasing $|m|$. Thus curves for $m = -4, \ldots, 4$ are shown. The spectral lines outside the gap correspond to scattering resonances with complex energies.}
\label{fig:fig_spectrum}
\end{figure}

States with $|m| > 0$, in turn, appear only when the defect is sufficiently large. In order to find the condition of supporting the state with some $m$ we notice that
$F_m(\epsilon, r_0)$ is monotonously decreasing function of energy while $\epsilon$ changes from $-\Delta$ to $\Delta$. Since $g_m(z \to 0) \sim (2m-1)!!/z^{m+1}$, we find that the energy of the $m$-th state is inside the gap, if
\begin{equation}
r_0 > R_m = \frac{v}{\Delta}\left(|m| - \frac{1}{2}\right).
\end{equation}
Conversely, for the given radius $r_0$ the number of bound defect states is given by $N = 4 + 8 \lceil r_0 \Delta/v + 1/2 \rceil$, where $4$ accounts for states from different valleys and with different spins at $m = 0$ and the second term accounts for states with $m > 1$, here $\lceil\ldots \rceil$ denotes taking the integer part and $8$ in addition to the spin and valley degeneracies accounts for the symmetry $m \to -m$. The dependence of energies of the defect states on the radius of the anti-dot is shown in Fig.~\ref{fig:fig_spectrum}.

The spatial electron distribution corresponding to the defect state is conveniently characterized by the probability density $\rho_{\tau, m}(r) = \braket{{\Phi}_{\tau, m}}{{\Phi}_{\tau, m}}$ and the vector of orientation of the (pseudo)spin coherent state $\mathbf{S}_{\tau, m} = \matrixel{{\Phi}_{\tau, m}}{ \bosi}{{\Phi}_{\tau, m}}/\rho_{\tau, m}$. As follows from Eq.~\eqref{eq:Phi_sc},
the pseudospin state is transversal, $S_r = 0$, with the spatial variation of the projection of $\mathbf{S}_{\tau, m}$ onto the $(\mathbf{e}_\phi, \mathbf{e}_z)$-plane depending on $m$.

In the simplest case $m=0$ the pseudospin remains in the plane of the layer, $S_{\tau,z}=0$. States with nonzero $m$ are characterized by out of the plane distribution of the pseudospin (for $r > r_0$). The angle of maximum deviation from the plane is
\begin{equation}\label{eq:ang_max}
\tan\left(\frac{\beta}{2}\right) = \tau
\frac{\sqrt{\Delta + \epsilon} - \sqrt{\Delta - \epsilon}}{\sqrt{\Delta + \epsilon} + \sqrt{\Delta - \epsilon}}.
\end{equation}
Thus for $\tau = 1$ the pseudospin ``sticks out" of the plane for $\epsilon > 0$ (that is for $m>0$) and has the negative projection on the $z$-axis for $\epsilon < 0$. For $\tau = -1$ the direction of the pseudospin is reversed.

\section{Optical Response}
\label{sec:optical}
\begin{figure*} 
	\begin{center}
		\includegraphics[width=7in]{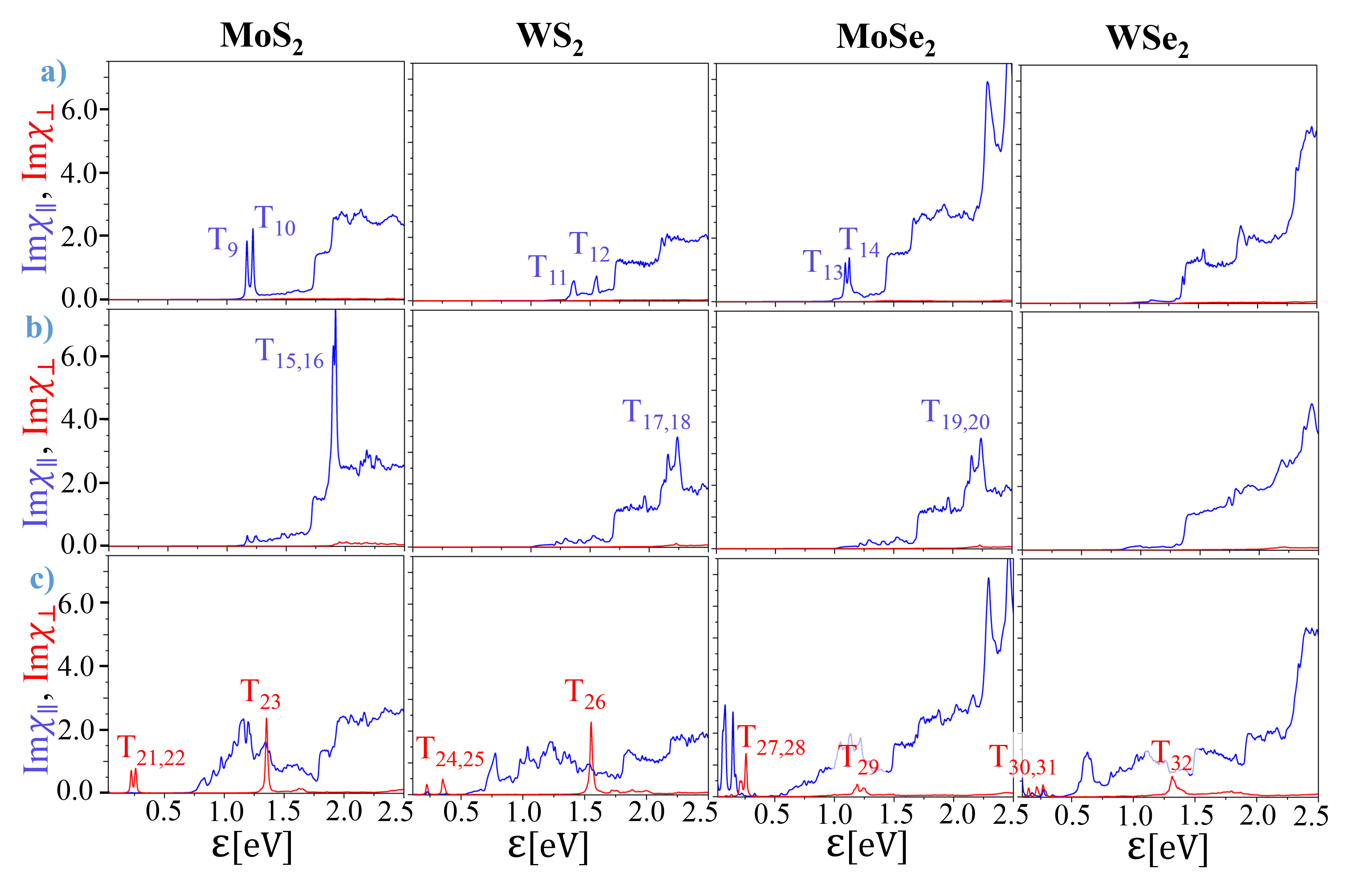}
	\end{center}
	\caption{Resonances of Im$\chi_{\parallel}(\omega)$ (blue) and Im$\chi_{\perp}(\omega)$ (red) $a)$ for X-vacancy, $b)$ X$_{2}$-vacancy, $c)$ M-vacancy for different TMDCs}
	\label{fig:OS_X_X2_M-vacancies}
\end{figure*}
The presence of LDS in the band structure gives rise to sharp peaks in the optical spectrum. \textcolor{black}{In Ref.~\onlinecite{Dielectric_function_measurement} the relative dielectric functions $\epsilon_r$ of various TMDCs have been measured experimentally, which in turn are related to the electric susceptibilities by the standard formula $\epsilon_r=1+\chi$.} In Fig.~\ref{fig:OS_X_X2_M-vacancies} results for both in plane $\chi_{\parallel}$ and out of plane $\chi_{\perp}$ components of the electric susceptibility are presented for different VDs in TMDCs. The electric susceptibility provides valuable insight into the optical selection rules for transitions between states across the Fermi level. We are interested in transitions involving states with energy near the gap edges or inside the gap. The electric susceptibility tensor is evaluated using the Kubo-Greenwood formula
\begin{equation}\label{eq:KGW}
\chi_{ij}(\omega)=-\frac{e^{2}\hbar^{4}}{m^{2}\epsilon_{0} V \omega^{2}}\sum_{p,q}\frac{f(E_{q})-f(E_{p})}{E_{pq}-\hbar \omega-i\hbar\Gamma}\pi_{pq}^{i}\pi_{qp}^{j},
\end{equation}  
where $\pi_{pq}^{j}=\matrixel{{\psi}_{p}}{x_{j}}{\psi_{q}}$ is the $j$th component of dipole matrix element between states $p$ and $q$, $V$ the volume, $f$ the Fermi function and $\Gamma$ is the broadening, which is set to be 0.01 eV. Appearance of states inside the band gap $E_{g}$ or close to the band edges leads to the resonances at single energy $E_{pq}=|E_{p}-E_{q}|$. The dipole matrix element $\pi_{pq}^{j}$ determines the strength of an optical transition and whether it is allowed or prohibited by symmetries.

\begin{figure}
\centering
\includegraphics[width=8cm]{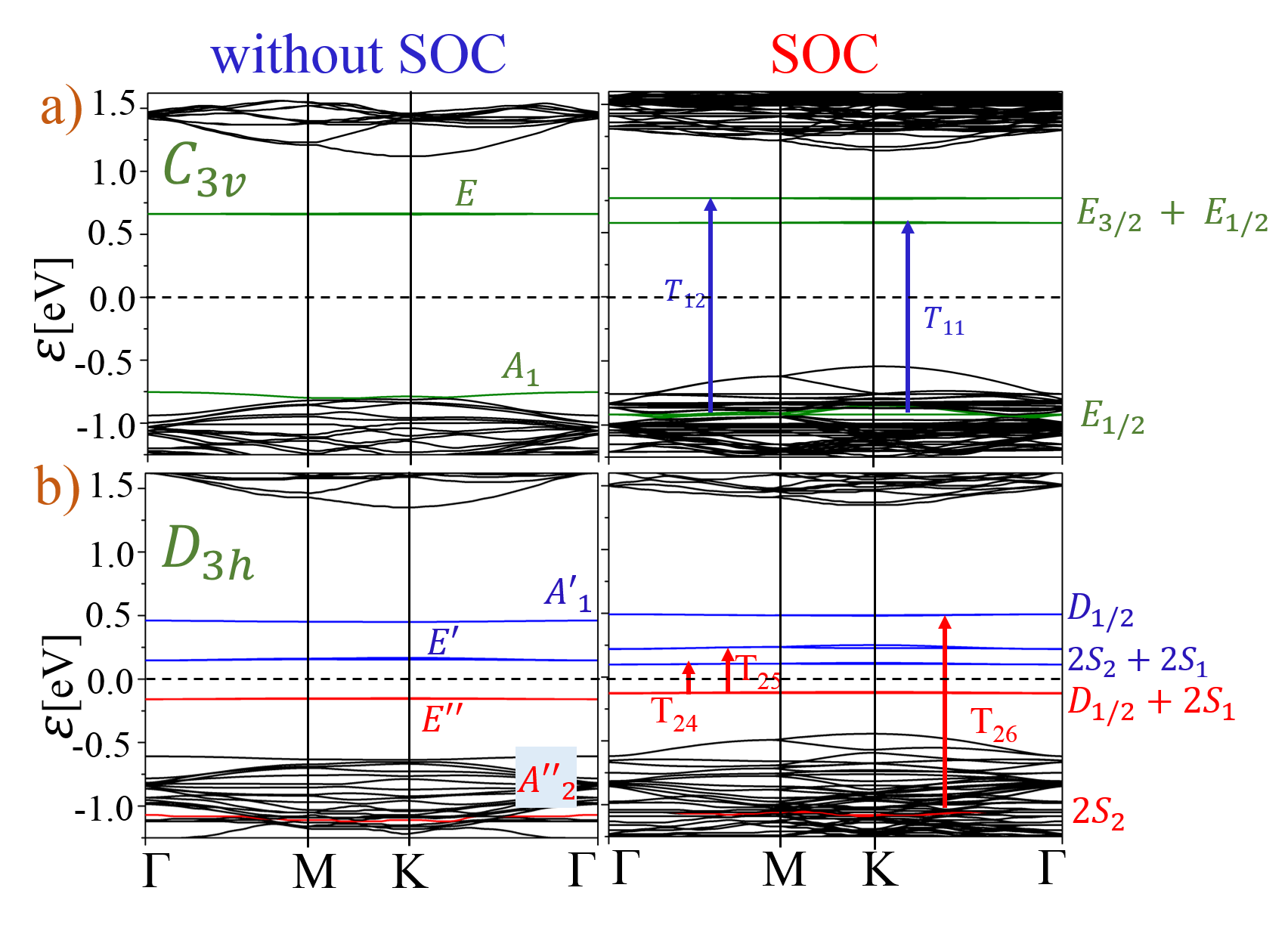}
\caption{Band structures without a) and with SOC b) for WS$_{2}$. The mapping of the LDS from the band structure without SOC to the band structure with SOC follows the mapping of the single group IRs to the double group IRs including spin, as explained in Table~\ref{table_D_D_3h}.}
\label{fig:single_Double_IR}
\end{figure}

When considering defects in a crystal, the LDS transform according to the IRs of the symmetry group of the crystal site in which the defect resides. While the translational symmetry of the crystal is broken, point group symmetries are partially or completely preserved. M and X$_{2}$ vacancies keep the $D_{3h}$ symmetry whereas the X-vacancy exhibits the lower $C_{3v}$ symmetry. The character tables for $C_{3v}$ and $D_{3h}$ with single and double group IRs  are shown in Table~\ref{table_C_3v} and Table~\ref{table_D_3h}, respectively. Table~\ref{table_D_D_3h} contains the decomposition of the direct products of the single group representations with the representation according to which the spin matrices transform, i.e. $E_{1/2}$ and $D_{1/2}$ respectively. In Fig:~\ref{fig:single_Double_IR} band structures for WS$_{2}$ with S and W vacancies are shown for with and without SOC. Following Ref.~\onlinecite{Selection_Rules_Dresselhaus, Selection_Rules_Birman}, in Fig:~\ref{fig:single_Double_IR} we show how the single group IRs in the absence of SOC can be mapped to the corresponding double group IRs in the presence of SOC. Note that due to the nature of the DFT calculation, the superlattice defined by the supercell introduces an artificial translational symmetry, which in some cases leads to artificial splittings. These can be typically recognized by systematically changing the size of the supercell.

The appearance of LDS inside the band gap leads to sharp resonances in $\chi_{\parallel}$ and $\chi_{\perp}$ at frequencies corresponding to the energy differences between LDS. However, not all transitions are allowed. Instead, several transitions are prohibited due to symmetry, i.e. when $\pi_{pq}^{j}$ does not transform according to the symmetric representation of the symmetry group of the superlattice. In the matrix element $\chi_{pq}^{j}$, the initial state $\psi_{p}$, the final state $\psi_{q}$, and the position operator $x_{j}$ transform according to the IRs $\Gamma(\psi_{p})$, $\Gamma(\psi_{q})$ and $\Gamma(x_{j})$, respectively. An electric dipole transition between two states is allowed if the direct product $\Gamma(\psi_{p})\otimes\Gamma(x_{j})\otimes\Gamma(\psi_{q})$ includes $\Gamma(I)$ in its decomposition in terms of a direct sum. $\Gamma(I)$ denotes the IR for the identity i.e., $A_{1}$ and $A_{1}'$ for $C_{3v}$ and $D_{3h}$, respectively. This is strictly related to the polarization of the radiation. One needs to consider separately the in plane and out of plane components of $\pi_{pq}^{j}$ because \textcolor{black}{they transform} according to different IRs of $C_{3v}$ and $D_{3h}$. The selection rules for electric dipole transitions for the double group IRs are summarized in Table~\ref{table_D_3h_C_3v}.

\begin{table}[tp]%
\caption{Character table of the group $C_{3v}$. $E$, $C_{3}$, $\sigma_{v}$ are the single group IRs and $E_{1/2}$, $E_{3/2}$ are the corrresponding double group IRs.}
\centering%
\begin{tabular}{|c| c| c| c|}\hline
                                                  $C_{3v}$         &    $E$    &    $C_{3}$    &       $3\sigma_{v}$\\ 
\hline
 $A_{1}$           & 1        & 1                      & 1                            \\
 $A_{2}$           & 1        & 1                      & $-1$                       \\
 $E$                    & 2        & $-1$                     & 0                           \\
\hline
 $E_{1/2}$                   & $2$   $-2$ &   $1$  $-1$                  & 0 0                \\
$E_{3/2}$         & $2$    $-2$ &   $-2$      $2$         & $0$        $0$        \\
\hline\end{tabular}
\label{table_C_3v} 
\end{table}

\begin{table}[tp]%
\caption{Character table of the group $D_{3h}$. $E$, $\sigma_{h}$, $2C_{3}$, $2S_{3}$, $3C_{2}$, and $\sigma_{v}$ are the single group IRs and $D_{1/2}$, $2S_{1}$, $2S_{2}$ are the corrresponding double group IRs.}
\centering%
\begin{tabular}{ |c |c |c |c |c |c |c |}\hline
                                                  $D_{3h}$                          &    $E$    &    $\sigma_{2}$    &    $2C_{3}$    &    $2S_{3}$    &    $3C_{2}$    &    $3\sigma_{v}$\\ 
\hline
        $A'_{1}$           & 1        & 1                      & 1                 & 1                                  &            1                          &  1                        \\
 $A'_{2}$           & 1        & 1                      & 1                 &  1                                &          -1                          & -1                        \\
 $A''_{1}$           & 1        & -1                     & 1                 &  -1                              &           1                          & -1                        \\
  $A''_{2}$           & 1        & -1                     & 1                 & -1                              &          -1                          &  1                         \\
 $E'$                    & 2        & 2                      & -1                  & -1                            &         0                          &  0                        \\
 $E''$                   & 2        & -2                      & -1                  & 1                            &         0                          &  0                        \\
\hline
                                                    $D_{1/2}$         & 2    -2 & 0                       & 1        -1        & $\sqrt{3}$    $-\sqrt{3}$                    &         0                          &  0                        \\
                                                    $2S_{1}$          & 2    -2       & 0                      & -2         2        & 0                    0             &            0                     &  0                        \\
                                                    $2S_{2}$          & 2   -2       & 0                      & 1         $-1$         & $-\sqrt{3}$    $\sqrt{3}$                &         0                          &  0                        \\
\hline\end{tabular}
\label{table_D_3h}
\end{table}

\begin{table}[h]
\begin{minipage}{80mm}
\begin{tabular}{|c |c |c |c |c}\hline
                                                  $\Gamma_{i}(C_{3v})$         &    $A_{1}$    &    $A_{2}$    &       $E$\\ 
\hline
 $\Gamma_{i}\times E_{1/2}$           & $E_{1/2}$        & $E_{1/2}$                      & $E_{3/2}+E_{1/2}$                            \\
\hline\end{tabular}

\end{minipage}
\begin{minipage}{80mm}

\begin{tabular}{|c |c |c |c |c |c |c|}\hline
                                                  $\Gamma_{i}(D_{3h})$         &    $A'_{1}$    &    $A'_{2}$    &    $A''_{1}$    &    $A''_{2}$    &    $E'$    &    $E''$\\
\hline
 $\Gamma_{i}\times D_{1/2}$           & $D_{1/2}$      & $D_{1/2}$  & $2S_{2}$  & $2S_{2}$  & $2S_{1}+2S_{2}$                      & $2S_{1}+D_{1/2}$                            \\
\hline\end{tabular}
\end{minipage}
\caption{Double group representations obtained from single group representaion for $C_{3v}$ and $D_{3h}$.}
\label{table_D_D_3h}
\end{table}

\begin{table}[h]
\begin{minipage}[b]{30mm}
\begin{tabular}{|c |c |c|}\hline
                                                  $C_{3v}$         &    $E_{1/2}$    &    $E_{3/2}$   \\
\hline
                                                  $E_{1/2}$        & $\sigma, \pi $       & $\sigma $     \\
\hline
                                                  $E_{3/2}$        & $\sigma $       & $\sigma, \pi$     \\
\hline\end{tabular}
\end{minipage}\begin{minipage}[b]{30mm}
\begin{tabular}{|c |c |c |c|}\hline
                                                  $D_{3h}$         &    $D_{1/2}$    &    $2S_{1}$ & $2S_{2}$   \\
\hline
                                                  $D_{1/2}$        &                         & $\sigma$      &  $\sigma,\pi$      \\
\hline
                                                  $2S_{1}$        & $\sigma$                  & $\pi$ &    $\sigma$ \\
\hline
				       $2S_{2}$        & $\sigma, \pi $       & $\sigma$ &     \\
\hline\end{tabular}
\end{minipage}
\caption{Electric Dipole selection rules in $C_{3v}$ and $D_{3h}$ symmetry. $\sigma$ represents in plane transitions while $\pi$ represents out of plane transitions.}
\label{table_D_3h_C_3v}
\end{table}

The presence of SOC couples the spin and orbital angular momenta, thereby requiring the consideration of the double group IRs. In our case, we need to consider $D_{1/2}\otimes C_{3v}$ and $D_{1/2}\otimes D_{3h}$, where $D_{1/2}$ is the 2-dimensional spin representation. The electromagnetic field couples to the orbital part of the state,  i.e. either to $\left|\zeta\right>$ or to $\left|\xi\right>$ [see Eq.~(\ref{Generalized_SOC})], polarizations induced by electromagnetic waves will not be changed in the presence of SOC. The role of the SOC is to lift some degeneracies, which gives rise to extra absorption peaks compared with the case without SOC.\cite{First_paper} In the susceptibility, these extra peaks lie close to the energies predicted by the susceptibility without SOC. It is important to take care when dealing with selection rules described by double groups because double groups may allow some transitions that are prohibited by the single groups. Then such selection rules must be discarded. \textcolor{black}{One such example is the $\pi$-transition for $X$-vacancy. In the abscence of SOC the $\pi$-transition is allowed only between states with the IRs $E$, i.e. $E\otimes A_{1}\otimes E=A_{1}\oplus A_{2} \oplus E$; this transition is not allowed since orbitals of the IR $E$ exist above the Fermi level only. In the presence of SOC the $\pi$-transition is allowed by the double group, i.e. $E_{1/2}\otimes A_{1}\otimes E_{3/2}=A_{1}\oplus A_{2} \oplus E$, but is not seen in the susceptibility including SOC. This can be understood as the artefact of double groups since $\pi$-transitions are prohibited in the abscence of SOC.} 

As a final note, as mentioned earlier, all samples are geometrically optimized before performing electronic calculations. Geometrical optimization may break certain symmetries and can affect certain selection rules or can result in concealing of some of the resonances.

\section{Conclusion}
In this paper we have provided numerical and analytical descriptions of electronic and optical properties of SL TMDCs in the presence of VDs. We have shown that the presence of LDS gives rise to sharp transitions both in $\chi{\parallel}$ and $\chi_{\perp}$. In order to understand these transitions, odd states need to be considered in addition to even states. A central result of our paper is that group theory can be used to derive strict selection rules for the optical transitions, which are in excellent agreement with the susceptibility calculated using the Kubo-Greenwood formula using the DFT orbitals. SOC induced splitting is observed in LDS and is seen to be larger for VDs in WX$_{2}$ than in MoX$_{2}$. Interestingly, our findings suggest magnetic properties of MoSe$_{2}$ in the presence of Mo vacancy, which may be enhanced  by increasing the density of defects. In order to provide a qualitative explanation of the existence of LDS, we performed analytical calculations based on the TBM and 2D Dirac formulation. All these results considerably improve the understanding of VDs in SL TMDCs and should benefit their potential applications in optoelectronic and nanoelectronic devices.
\\

\section{Acknowledgments}
We acknowledge support provided by NSF CCF-1514089, the Airforce Summer Faculty Fellowship 2016, and AFOSR LRIR 15RY159COR. We thank Kavir Dass, Shin Mou, and Xiaodong Xu for useful discussions.

\bibliographystyle{apsrev4-1}

\end{document}